\documentclass[prl,aps,twocolumn,superscriptaddress]{revtex4}
\usepackage{graphicx,amssymb,amsmath,color,psfrag}
\usepackage{amsthm}
\usepackage{amsfonts}
\usepackage{algorithmic}
\usepackage{enumerate}
\usepackage{latexsym}

\begin{document}

\title{Ground state pairing correlations in the $S_4$ symmetric microscopic model for iron-based superconductors}
\author{Yang Wu}
\affiliation{Department of Physics, Beijing Normal University,
Beijing 100875, China\\}
\author{Guangkun Liu}
\affiliation{Department of Physics, Beijing Normal University,
Beijing 100875, China\\}

\author{Tianxing Ma}
\email{txma@bnu.edu.cn}
\affiliation{Department of Physics, Beijing Normal University,
Beijing 100875, China\\}

\affiliation{Beijing Computational Science Research Center,
Beijing 100084, China}

%
%
\date{\today}

\begin{abstract}
We present the ground state pairing correlations in the $S_4$ symmetric microscopic model for iron-based superconductors,
computed with the constrained-path Monte Carlo method.
For various electron fillings and interaction strengths, we find that the $s_{xy}$ pairing dominate over other pairing correlations and are positive when the pair separation exceeds several lattice constants, whatever for iron pnictides and iron chlcogenides. These ground state properties, especially the long range part pairing correlations re-conform the previous finite temperature results published in Phys. Rev. Lett. 110, 107002(2013). We further our study by including the nearest neighbour interaction $V$ and it is found that the $s_{xy}$ pairing correlation is slightly suppressed by the increasing $V$.
\end{abstract}

\maketitle

\section{Introduction}

Iron-based superconductors are the major field in superconductivity now\cite{Hosono,ChenXH,nlwang,ChenXL}.
One of today's major challenge in the study of iron-based superconductors is how to obtain an unified microscopic understanding of the different families of these materials, in particular, iron-pnicitides and iron-chalcogenides, which distinguish themselves from each other with distinct Fermi surface topologies\cite{Wang_122Se, Zhang_122Se,Mou_122Se}. Theoretical studies based on models with complicated multi-d orbital band structures\cite{john,hirschfeld,Dongj2008,Mazin2008,Kuroki2011, WangF, thomale1, thomale2, chubukov,zlako,Arita2009}, lack of a support from  more fundamental microscopic electronic physics\cite{Maiti2011,seo2008,Si2008,Fang2008nematic,Ma2008lu,hu1,hu4,luxl,berg}.
Recently, it has been shown that the underlining electronic structure in iron-based superconductors,
which is responsible for superconductivity at low energy, is essentially governed by a two-orbital model with a $S_4$ symmetry.
The two orbital model includes two nearly degenerated single-orbital parts that can be mapped to each other under the $S_4$ transformation.
This electronic structure  stems from  the fact that the dynamics of  $d_{xz}$ and $d_{yz}$ orbitals are divided into two groups that are separately coupled to the top and bottom As(Se)
planes in a single Fe-(As)Se trilayer structure. The two groups can thus be treated as an $S_4$ iso-spin.
The dressing of other orbitals in the $d_{xz}$ and $d_{yz}$ orbitals can not alter the symmetry characters. 
Despite the simplicity of this description, the model not only gives very good quantitative
agreement with the band structure but also supply a uniform model to mimic
different iron-based material classes, especially for the iron-pnictides and
iron-chalcogenides\cite{Hu2012s4}.

Some of us have performed a finite temperature determinant quantum Monte Carlo (DQMC) study of the
pairing correlation in the $S_4$ symmetric microscopic model on lattices mainly with $8^2$ 
sites.
It is found that the pairing with an extensive $s$-wave symmetry robustly dominates over other pairings at low temperature in reasonable parameter
region regardless of the change of Fermi surface topologies.
The pairing susceptibility, the effective pairing interaction and the $(\pi,0)$ antiferromagnetic
correlation  strongly increase as the on-site Coulomb
interaction increases,
and these non-biased
numerical results provide a possible unified understanding of superconducting
mechanism in iron-pnictides and iron-chalcogenides\cite{Ma2013}.

Numerical approaches like DQMC, however, had its own difficulties, typically being limited to small sizes, high temperature, and
experience the infamous fermion sign problem, which cases exponential growth in the variance of the computed results and
hence an exponential growth in computer time as the lattice size is increased and the temperature is lowered\cite{Blankenbecler1981,MaReview2011}.
In general, to determine which pairing symmetry is dominant by numerical
calculation for finite size models, we have to look at the long-range
part of the pair-correlation function at zero temperature, and it seems to be dangerous to
extrapolate the long-range behavior of the pair-correlation function
from a lattice with $8^2$ sites.
In order to shed light on this critical issue, it is important to discuss the results obtained from the constrained path Monte Carlo (CPMC)\cite{Zhangcpmc} on larger lattice. In a variety of benchmarking calculations, the CPMC method has yielded very accurate results of
the ground state energy and many other ground state observables for large system\cite{Zhangcpmc}.
In the CPMC method, the ground-state wave
function $|\Psi_0>$ is projected from an initial wave function $|\Psi_T>$  by a branching
random walk in an over-complete space of constrained Slater
determinants $|\phi>$, which have positive overlaps with a known trial wave
function. In such a space, we can write $|\Psi_0>=\Sigma_{}\phi \chi(\phi)|\phi>$, where  $\chi(\phi)>0$. The random walk produces an ensemble of
 $|\phi>$, called random walkers, which represent $|\Psi_0>$ in the sense that their distribution is a Monte Carlo sampling of $\chi(\phi)$. The resulting method is free of $any$ decay of the signal-to-noise ratio.
For more technique details we refer to
Refs.~\cite{Zhangcpmc,Huangcpmc}.

In this paper, we report the ground state results in the $S_4$ symmetric microscopic model for various electron fillings, interaction strength by using CPMC method.
The simulations were mainly performed on a $12^2$ lattices, and compared to the paring correlation on an $8^2$, a $16^2$ and a $20^2$ lattices. All the lattices are  with periodic boundary conditions.
Our unbiased numerical calculation shows that the ground state $s_{xy}$ pairing dominate over other pairing correlations. The $s_{xy}$ pairing correlations is  positive when the pair separation exceeds several lattice constants, whatever for iron pnictides and iron chlcogenides. These ground state properties, especially the long range part pairing correlations re-conform our previous finite temperature results with DQMC method\cite{Ma2013}. We further our study by including the nearest-neighbor interaction $V$. It is found that the $s_{xy}$ pairing correlation is slightly suppressed by the increasing $V$.

\section{Model and Results}

The minimum extended Hubbard model for a single $S_4$ iso-spin component in the iron-square lattice is described by\cite{Hu2012s4,Ma2013},
\begin{eqnarray}
&&H=t_{1}\sum_{\mathbf{i}\eta \sigma }(a_{\mathbf{i}\sigma }^{\dag }b_{%
\mathbf{i}+\eta \sigma }+h.c.) \notag \\
&&+ t_{2}[\sum_{\mathbf{i}\sigma }a_{\mathbf{i}\sigma }^{\dag }a_{\mathbf{i}%
\pm (\hat{x}+\hat{y}),\sigma }+\sum_{\mathbf{i}\sigma }b_{\mathbf{i}\sigma }^{\dag }b_{\mathbf{i}\pm (\hat{x}-\hat{y})\sigma
}] \notag \\
&&+t_{2}^{\prime }[\sum_{\mathbf{i}\sigma }a_{\mathbf{i}\sigma }^{\dag }a_{%
\mathbf{i\pm }(\hat{x}-\hat{y})\sigma }+\sum_{\mathbf{i}\sigma }b_{\mathbf{i}\sigma }^{\dag }b_{\mathbf{i\pm (}%
\hat{x}+\hat{y})\sigma }]\notag\\
&&+U\sum_{\mathbf{i}}(n_{a\mathbf{i}\uparrow}n_{a\mathbf{i}\downarrow} + n_{b\mathbf{i}\uparrow}n_{b\mathbf{i}\downarrow})+V\sum_{\mathbf{i}\eta \sigma }(a_{\mathbf{i}\sigma }^{\dag }b_{%
\mathbf{i}+\eta \sigma }+h.c.) \notag \\
&&+\mu\sum_{\mathbf{i}\sigma}(n_{a\mathbf{i}\sigma }+n_{b\mathbf{i}\sigma })
\end{eqnarray}
Here, $a_{i\sigma}$ ($a_{i\sigma}^{\dag}$) annihilates (creates) electrons
at site $\mathbf{R}_i$ with spin $\sigma$ ($\sigma$=$\uparrow,\downarrow$)
on sublattice A, $b_{i\sigma}$ ($b_{i\sigma}^{\dag}$) annihilates (creates)
electrons at the site $\mathbf{R}_i$ with spin $\sigma$
($\sigma$=$\uparrow,\downarrow$) on sublattice B,
$n_{ai\sigma}=a_{i\sigma}^{\dagger}a_{i\sigma}$,
$n_{bi\sigma}=b_{i\sigma}^{\dagger}b_{i\sigma}$, $\eta=(\pm \hat{x},0)$ and $(0,\pm \hat{y})$.
$U$ and $V$ denote the on-site
Hubbard interaction and NN interaction, respectively.
In  the above model, for simplicity and clarity, we only keep a minimum set of parameters which include three key shortest hopping parameters that are responsible for the physical picture revealed by the $S_4$ symmetry\cite{Hu2012s4}.
The selection of parameters in following studies does capture the essential physics of typical cases for iron-pnictides\cite{hding,hding2,lxyang,ludh,chenf} and iron-chalcogenides\cite{Wang_122Se, Zhang_122Se,Mou_122Se}, as that shown in Ref.\cite{Ma2013}.

\begin{figure}[tbp]
\includegraphics[scale=0.45]{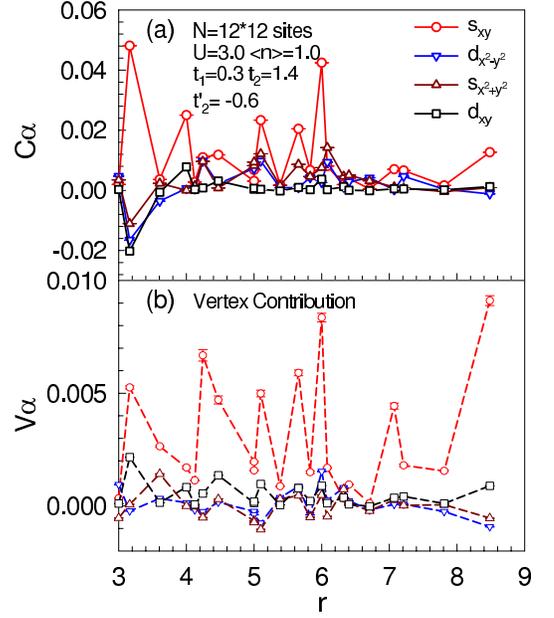}
\caption{(Color online) (a) Pairing correlation $C_{\alpha}$ as a
function of distance for different pairing symmetries on
the $12^2$ lattice with
$t_{1}=0.3, t_{2}=1.4, t'_2=-0.6$ (a typical case for iron-pnictides\cite{hding,hding2,lxyang,ludh,chenf}). (b) The vertex contribution  $\overline{V_{\alpha}}$ with the same parameters.} \label{Fig:Symmetry1}
\end{figure}

The pairing correlation function we computed is
\begin{eqnarray}
C_{\alpha }({\bf{r=R_{i}-R_{j}}})=\langle \Delta _{\alpha }^{\dagger }
(i)\Delta _{\alpha }^{\phantom{\dagger}}(j)\rangle ,
\end{eqnarray}
where $\alpha$ stands for the pairing symmetry. And the corresponding order
parameter $\Delta_{\alpha }^{\dagger }(i)$\ is defined as
\begin{eqnarray}
\Delta_{\alpha }^{\dagger }(i)\ =\sum_{l}f_{\alpha}^{\dagger}
(\delta_{l})(a_{{i}\uparrow }b_{{i+\delta_{l}}\downarrow }-
a_{{i}\downarrow}b_{{i+\delta_{l}}\uparrow })^{\dagger},
\end{eqnarray}
with $f_{\alpha}(\bf{\delta}_{l})$ being the form factor of pairing
function, and the vectors $\bf{\delta_{l}}$($\bf{\delta_{l}^{\prime}}$
denote the nearest neighbour intersublattice connections (the next nearest neighbour inner sublattice
connections), where $l=1,2,3,4$ denoting the four different direction.
As that shown in Ref.\cite{Ma2013}, we focus on four kinds of pairing form, where
\begin{eqnarray}
\text{$d_{x^2-y^2}$-wave} &\text{:}&\ f_{d_{x^2-y^2}}(\delta_{l})=1
~(\delta_{l}=(\pm\hat{x},0)) \notag \\
\text{and} &\text{:}&\ f_{d_{x^2-y^2}}(\delta_{l})=-1
~(\delta_{l}=(0,\pm\hat{y})) \notag \\
\text{$d_{xy}$-wave} &\text{:}&\ f_{d_{xy}}(\delta_{l}^{\prime})=1
~(\delta_{l}^{\prime}=\pm(\hat{x},\hat{y})) \notag \\
\text{and} &\text{:}&\ f_{d_{xy}}(\delta_{l}^{\prime})=-1
~(\delta_{l}^{\prime}=\pm(\hat{x},\hat{-y}))\notag \\
\text{$s_{x^2+y^2}$-wave}   &\text{:}&\ f_{s_{xy}}(\delta_{l})=1,~l=1,2,3,4 \notag \\
\text{$s_{xy}$-wave} &\text{:}&\ f_{s_{xy}}(\delta_{l}^{\prime})=1,~l=1,2,3,4
\end{eqnarray}
To facilitate contact with prior simulations, we also examined the vertex contributions to the correlations defined by
\begin{eqnarray}
V_{\alpha}(\bf{r})=C_{\alpha}(\bf{r})-\overline{C_{\alpha}}(\bf{r})  \label{Vertex}
\end{eqnarray}
where $\overline{C_{\alpha}}(\bf{r})$ is shorthand notation for the uncorrelated pairing correlation. For each term in
${C_{\alpha}}(\bf{r})$ like $\langle a^{\dagger}_\uparrow a_\uparrow a^{\dagger }_\downarrow a_\downarrow \rangle$,
it has a term like$\langle a^{\dagger }_\uparrow a_\uparrow \rangle \langle a^{\dagger}_\downarrow a_\downarrow \rangle$.


\begin{figure}[tbp]
\includegraphics[scale=0.45]{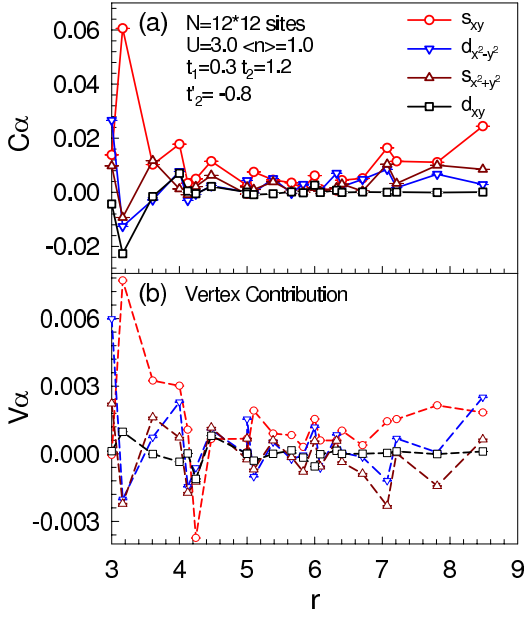}
\caption{(Color online) (a) Pairing correlation $C_{\alpha}$ as a
function of distance for different pairing symmetries on
the $12^2$ lattice with $t_{1}=0.3, t_{2}=1.2, t'_2=-0.8$. (b) The vertex contribution  $\overline{V_{\alpha}}$ with the same parameters.} \label{Fig:Symmetry2}
\end{figure}

In Fig.~\ref{Fig:Symmetry1} (a), we compare the long-range part of pairing
correlations with different pairing symmetries on the $12^2$ lattices at $t_{1}=0.3, t_{2}=1.4, t'_2=-0.6$, which is a typical case for iron-pnictides\cite{hding,hding2,lxyang,ludh,chenf}.
Here, the electron filling is $<n>=1.0$, which corresponds to a closed shell case with $N_\uparrow$=$N_\downarrow$=72. The simulations are performed at $U=3.0$.
One can readily see that $C_{s_{xy}}(r)$ (solid red line) is larger than pairing correlations with any other pairing symmetry for almost all long-range distances between electron pairs.
With the same set of parameters as that of Fig.~\ref{Fig:Symmetry1} (a), Fig.~\ref{Fig:Symmetry1} (b) shows the vertex contribution defined in Eq. \ref{Vertex}. Obviously,
the vertex contribution of $s_{xy}$ ( dash red line) pairing symmetry dominate that of any other pairing forms. The vertex contribution of $s_{xy}$ pairing symmetry is a finite value over the long range part, while vertex contributions of $d_{xy}$, $s_{x^2+y^2}$ and $d_{x^2-y^2}$ are simply fluctuating around zero.
In the numerical results, the ratio of the statistical error to the pairing correlation $C_{\alpha}$ is no more than 0.5 percent,
and most of the error bars are almost within the symbols.
The ratio of the statistical error to the vertex contribution $\overline{V_{\alpha}}$ is no more than 3 percent. 
This remark applies to all the following figures. 

Fig.~\ref{Fig:Symmetry2} shows the long-range part of pairing correlations with different pairing symmetries on the $12^2$ lattice at $t_{1}=0.3, t_{2}=1.2, t'_2=-0.8$.
With this set of parameters, the system shows no hole packet\cite{Ma2013}. Again we see that, both the long range part pairing correlation and the vertex contribution indicates that the $s_{xy}$ type dominates over that of other pairing forms. Thus, the behavior of long-range part pairing correlation re-enforces our findings on pairing susceptibility of an $8^2$ lattice in Ref.\cite{Ma2013}.

\begin{figure}[tbp]
\includegraphics[scale=0.45]{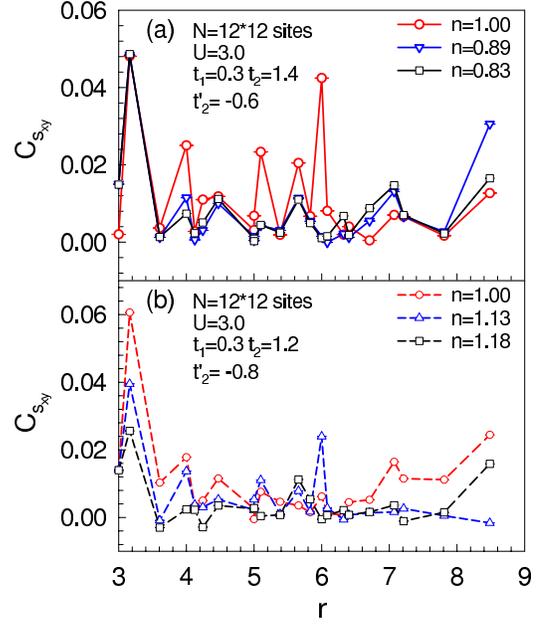}
\caption{(Color online) (a) Pairing correlation $C_{\alpha}$ as a
function of distance on the $12^2$ lattice with
$t_{1}=0.3, t_{2}=1.4, t'_2=-0.6$ for $<n>$=1.00, $<n>$=0.89 and $<n>$=0.83.
(b) Pairing correlation $C_{\alpha}$ as a
function of distance on the $12^2$ lattice with
$t_{1}=0.3, t_{2}=1.2, t'_2=-0.8$ for $<n>$=1.00, $<n>$=1.13 and $<n>$=1.18. } \label{Fig:Filling}
\end{figure}

\begin{figure}[tbp]
\includegraphics[scale=0.45]{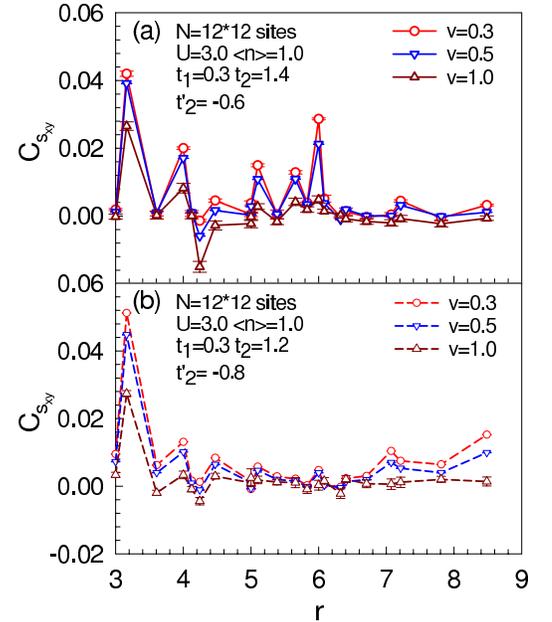}
\caption{(Color online) (a) Pairing correlation $C_{\alpha}$ as a
function of distance at the nearest neighbour interaction $V=0.3, 0.5$ and $1.0$ on the $12^2$ lattice with
$t_{1}=0.3, t_{2}=1.4, t'_2=-0.6$. (b) The same with (a) but at $t_{1}=0.3, t_{2}=1.2, t'_2=-0.8$.} \label{Fig:V}
\end{figure}

In Fig.~\ref{Fig:Filling}, we address the question of what happens to those ``long-range" correlations if the system is doped away from
half filling. In Fig.~\ref{Fig:Filling} (a), for a closed shell case with electron filling
$<n>=0.83$ ( $N_\uparrow$=$N_\downarrow$=60),  $<n>=0.89$ ( $N_\uparrow$=$N_\downarrow$=64) and
  $<n>=1.00$ ( $N_\uparrow$=$N_\downarrow$=72), we show the CPMC results of $s_{xy}$ pairing correlation
  for $U=3.0$ and $t_{1}=0.3, t_{2}=1.4, t'_2=-0.6$. 
Fig.~\ref{Fig:Filling} (b) shows results of $s_{xy}$ pairing correlation for $t_{1}=0.3, t_{2}=1.2, t'_2=-0.8$ at $<n>=1.00$, $<n>=1.13$ ( $N_\uparrow$=$N_\downarrow$=81) and  $<n>=1.18$ ($N_\uparrow$=$N_\downarrow$=85).
  We notice that, whatever for system with or without hole packet, the pairing correlations decrease as the system is doped away from half filled case.

We have also studied the effect of nearest neighbour interaction on the pairing correlation at a fix $U=3.0$.
In Fig. \ref{Fig:V}, the pairing correlations for $s_{xy}$ pairing symmetries are displayed as a function of distance
on the $12^2$ lattice with different nearest neighbour interaction $V's$.
For both systems with and without hole packet, we notice that the $s_{xy}$ pairing correlation is suppressed by the repulsive nearest neighbour interaction $V$.

\begin{figure}[tbp]
\includegraphics[scale=0.45]{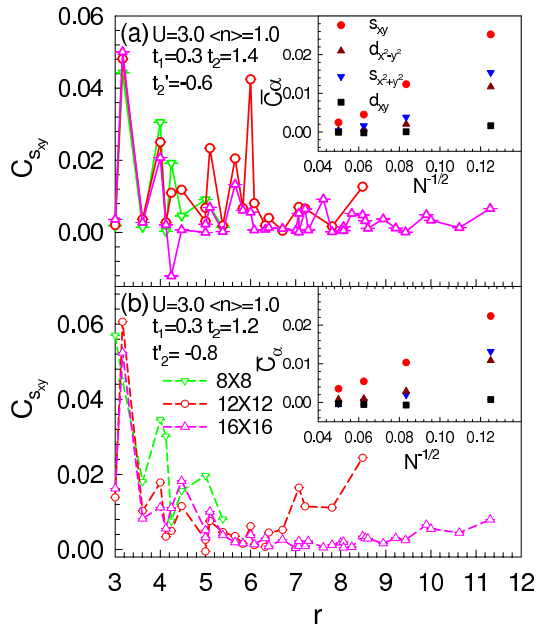}
\caption{(Color online) (a) Pairing correlation $C_{s_{xy}}$ as a
function of distance on an $8^2$, a $12^2$, and a $16^2$ lattices with
$t_{1}=0.3, t_{2}=1.4, t'_2=-0.6$.
(b) The same with (a) but at $t_{1}=0.3, t_{2}=1.2, t'_2=-0.8$; Inset: the average of
the long-range pairing correlation $\overline{C}_{\alpha}$ vs
$\frac{1}{\sqrt{N}}$ at half filling.} \label{Fig:Lattice}
\end{figure}

Finally, we compare the pairing correlation on an $8^2$ ( green triangle down ), a $12^2$ ( red circle ), and a $16^2$ ( pink triangle up ) lattices in Fig. \ref{Fig:Lattice} to exclude the size effect.
Fig.5 (a) shows the pairing correlation with $s_{xy}$ symmetry for $t_{1}=0.3, t_{2}=1.4, t'_2=-0.6$, and Fig.5 (b) shows the pairing correlation with $s_{xy}$ symmetry for $t_{1}=0.3, t_{2}=1.2, t'_2=-0.8$.
In the inset of Fig. \ref{Fig:Lattice}, we examine the evolution of $C_{\alpha}$
with increasing the lattice size up to $20^2$. The average of long-range pairing
correlation,
$\overline{C}_{\alpha}$=$\frac{1}{\sqrt{N'}}\sum_{r\geq3}C_{\alpha}(r)$,
where $N'$ is the number of electron pairs with $r\geq3$, is plotted
as a function of $\frac{1}{\sqrt{N}}$ at half filling. It is clear to see that  $\overline{C}_{s_{xy}}$ (red circle) is larger than the average of long-range pairing correlations with any other pairing symmetry for whichever lattice size we investigate.

\section{CONCLUSIONS}
In summary, our unbiased numerical results show that the $s_{xy}$ pairing dominate in the ground state
of the $S_4$ model, as we illustrated in previous study. And such a domination is robust
in a wide range of physical region. We also find that the nearest neighbour interaction slightly suppressed
the pairing correlation.
The consistent behaviours of our results on different clusters suggest that
$S_4$ model captures the essence of iron-based superconductors.

{\it Acknowledgement:} 
This work is supported by NSFCs (Grant. No. 11104014, No. 11374034 and No. 11334012),
Research Fund for the Doctoral Program of Higher Education of China
20110003120007, SRF for ROCS (SEM).


\end{document}